# Transient Stability evaluation of a Solar PV Integrated Industrial Microgrid

Ferdous Irtiaz Khan, Tareq Aziz
Department of Electrical and Electronic Engineering
Ahsanullah University of Science & Technology
Dhaka, Bangladesh
Email: firtiaz.04@gmail.com

*Abstract*—Whenever stability of a system is concerned, transient stability is given the highest priority. For analysis purpose transient stability is checked with the help of Critical Clearing Time (CCT). With the continuous growth of Distributed Generation (DG) concept, the need of a stable industrial microgrid is a necessity. This paper studies the effect of Solar PV as DG unit on CCT during islanded operation of an industrial microgrid system. In primary analysis a single centralized Solar PV unit along with the Synchronous Generator (SG) is used. Then multiple PV units distributed at random locations are used. In both cases the share by Solar PV is gradually increased and the effect on the length of CCT is observed. In the first case the CCT is found to increase with the inclusion of Solar PV and in the latter one CCT drops as multiple units are used.

## I. INTRODUCTION

Stability had always been a major concern for every power system. During analysis of stability different forms are studied like transient stability, dynamic stability and steady state stability. Among these, transient stability is given the utmost importance as it refers to the stability after major disturbances like faults or loss of generating units, etc. During the study of transient stability, the factors that are mainly exploited are Critical Clearing Time (CCT), location of fault, impact of load and angular swing of the rotor [1-3]. CCT is considered an important parameter for the evaluation of transient stability [4]. It is the maximum time that the fault can persist without the loss of synchronism of the system.

Modern Power system contains a large number of Distributed Generators (DGs) in the system and a good part of it is contributed by renewable sources. With renewable sources certain challenges are faced. As discussed by Muruganantham in [5] and Zheng in [6], due to inclusion of renewable sources voltage irregularities, system instability and decline of reliability is often faced. It often encounters generation interruption and lack of technology standard [7-9]. According to Longatt in [10] & [11] the electronic converter-based systems (Solar PV, Wind etc.) lowers the share of the synchronous machines and hence cause a reduction of overall system inertia. One of the major challenges that are encountered is to maintain transient stability.

Industrial Microgrids are complicated system with presence of large number of induction motors. In consideration of microgrids, industrial microgrids are the most challenging ones [12]. This paper deals with the impact that Solar PV as a generating unit possess in an industrial microgrid by analyzing the CCT for different scenarios. At first the CCT is observed by using a single Solar PV unit along with the existing Synchronous Generator (SG). The share by the PV unit is gradually increased and the difference is analyzed. Then the same procedure is followed but with the use of multiple Solar PV units placed at random locations in the system. For analysis purpose simulation is performed in DigSilent Powerfactory software on IEEE Std 399 system which is a built in 43 bus system.

The paper is arranged as follow: the introduction is already given in section I. In section II some basic concepts about transient stability and CCT are discussed. The solar PV model is provided in Section III. Section IV is the Results section where the simulation results for the analysis of Solar PV impact on CCT are shown. Finally, in section V a conclusion based on the obtained results is made.

## II. TRANSIENT STABILITY AND CCT

Transient stability is generally referred to as the ability of the system to regain synchronism after major disturbances like loss of generator, line switching, large load changes or faults. In [1] it is defined as the stability which maintains the synchronizing torque, also defined as first swing stability. According to [13] it is the scope of a system to retain a constant operating point after being subjected to disturbance that changes the system's topology. To accurately analyze transient stability numerical integration methods, also known as time domain methods, and graphical methods called Equal Area Criterion (EAC) are generally used. Under time domain analysis the robustness of the system is attained by observing the Critical Clearing Time (CCT) of the system [14].

There is a critical angle for clearing the fault so that the condition for equal area criterion is maintained. This angle is known as critical clearing angle and the corresponding time for clearing the fault is called CCT. Generally speaking, it is the maximum time needed to clear the fault such that the system does not loose synchronism after the fault is cleared. In terms of equation, CCT is derived from Critical Clearing Angle (CCA) using EAC. The equations are shown in (1) & (2)

$$\delta_{cr} = \frac{\omega_s P_m}{4H} t_{cr}^2 + \delta_0 \qquad (1)$$

$$t_{cr} = \sqrt{\frac{4H(\delta_{cr} - \delta_0)}{\omega_s P_m}} \qquad (2)$$

Where,
*H = Inertia Constant of machine*
*$\delta_{cr}$ = Critical Clearing angle*
*$\delta_0$ = Rotor angle when generator is at synchronous speed*
*$\omega_s$ = synchronous speed of machine in rad/sec*
*$P_m$ = Mechanical Power*
*$t_{cr}$ = Critcal Clearing Time*

From equation (2) it is visible that CCT mainly depends on factors such as CCA, rotor angle, synchronous speed, inertia of the machine and input mechanical power. These parameters on the other hand are affected by the type of machine, fault location, amount of load present, intensity of the fault current, etc. [3][15]

The concept of distributed generation (DG) is a widely accepted phenomena for modern power system. As discussed in [16] with distributed generation large power systems obtain back up supply, it increases efficiency, improves power quality and reliability, reduces line loss, transmission and distribution investment is lowered. In [17] it was found that distributed generation in a limited bus system makes the grid more stable by increasing the CCT but at the same time it may increase fault level and might result in lowering of bus voltage. Kuang in [16] further discussed that with the introduction of distributed generation, system topology is hampered due to the increment in the number of electronic interfaced devices, inductors and capacitors. Thus, this leads to an uneven stability of the entire system. As distributed source, wind turbines and solar PV are commonly used. In [18] & [19] it was mentioned that wind turbines can reduce system stability through high transient current, rotor overvoltage, lowering the effective inertia of the power system and thus hampering the frequency of the system. According to [20] a large variation in frequency and voltage may occur when distributed generation goes to islanded mode due to mismatch between generation and load present.

To improve the stability of a system different approaches are followed. Using Battery Energy Storage System (BESS) and STATCOM transient stability and in some cases CCT improvement is suggested in [21-23]. In [24] analysis proved that solar PV have a major impact in transient stability of the system and from [25] we can be assured that solar PV have a positive impact in transient stability. By ZhiWei in [26] CCT can be improved using Voltage Source Converter HVDC. As suggested by Alaboudy in [27], microgrid with 100% SG has lower CCT than that of system which combines SG and inverter interfaced DG. This paper deals with an Industrial microgrid and the effect that Solar PV share have on the length of the CCT. It also examines whether multiple Solar PVs affect CCT instead of a single large unit.

### III. MODELLING OF SOLAR PV

Figure 1 shows the block diagram of a single stage solar PV system. A solar PV mainly consists of array of PV panel, a voltage source converter (VSC) and a three-phase interface reactor [28].

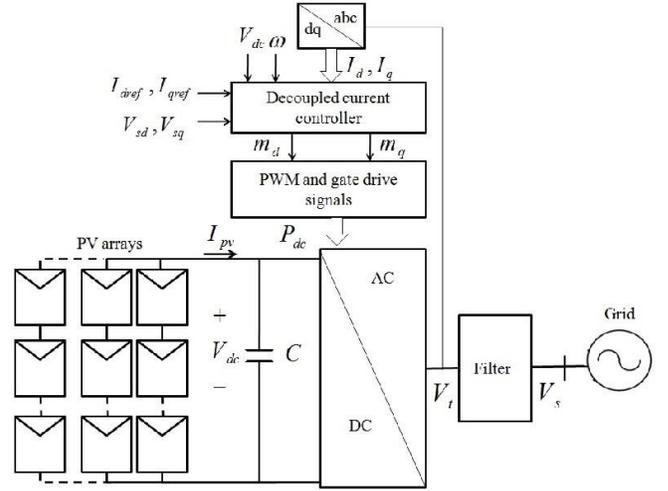

Fig. 1: Block diagram of single stage solar PV system [28]

In figure 1 the following terms are defined as,

$$I_{pv} = n_p I_{sc} - n_p I_{rs}\left[exp\frac{q}{kTA}\frac{V_{dc}}{n_s} - 1\right] \quad (3)$$

where,
*$I_{pv}$ = PV array current*
*$V_{dc}$ = PV array voltage*
*$n_p$ = number of parallel modules in PV array*
*$n_s$ = number of cells in series*
*$I_{sc}$ = short circuit current of each string*
*$I_{rs}$ = Cell reverse saturation current*
*k = Boltzman constant*
*q = unit electric charge*
*T = cell temperature*

The short circuit current is a linear function of Solar irradiation level. The power delivered by PV would be:

$$P_{pv} = I_{pv}V_{dc} \quad (4)$$

$$P_{pv} = n_p I_{sc} V_{dc} - n_p I_{rs} V_{dc}\left[exp\frac{q}{kTA}\frac{V_{dc}}{n_s} - 1\right] \quad (5)$$

From equation (5) we can see that $P_{pv}$ is a function of $V_{dc}$ and thus by controlling $V_{dc}$, $P_{pv}$ can be maximized. Based on the principle of power balance, the dynamics of DC link voltage can be described as

$$\frac{C}{2}\frac{dV_{dc}^2}{dt} = P_{pv} - P_{dc} \quad (6)$$

where $P_{dc}$ is the power delivered to VSC.

## IV. RESULTS AND ANALYSIS

### A. Test System & Simulation Platform

The test system under consideration is the IEEE Std 399-1997 system which is a 43-bus system and is shown in Fig 2. The system is considered an industrial system as it contains large number of motors (28 induction motors and 1 synchronous motor). Initially it is connected to the utility grid but later when a short circuit fault occurs at the Utility grid bus, it is disconnected from the system and a SG acting as DG which is connected at bus-4 remains connected to the microgrid section and CCT analyzed. Swing equation is mainly considered because in every analysis synchronous generator is present and as it dictates rotor dynamics. During every simulation the SG is kept connected at bus 4. In a similar fashion Solar PV, along with the existing SG, is also used and gradually its share increased to observe the change in the length of CCT. Further analysis is made using multiple Solar PV at random locations rather than a single centralized PV unit and CCT is observed.

As simulation tool DIgSILENT PowerFactory 15.1.7 has been used to observe voltage and maximum SG rotor angle deviation in order to analyze the change in CCT of the system.

Two different analyses are made:

Analysis 1: The microgrid sections CCT is analyzed using SG and gradually increasing share of a single centralized solar PV.

Analysis 2: The microgrid sections CCT is analyzed using SG and gradually increasing share of multiple solar PVs located at random buses.

### B. With SG and a single Solar PV unit

A short circuit fault is given at bus 4 at 2 seconds and then cleared at a later time. Bus 1 & 2 are disconnected immediately after the fault so that Utility grid is disconnected from the load part and a microgrid section is formed. Initially only SG remains connected at bus 4 and the fault clearing time gradually increased to calculate the CCT. The plots of voltage and maximum rotor angle deviation is given in figure (3) and figure (4). It can be seen that beyond 6.2s system fails to recover as voltage doesn't return to nominal value and the maximum rotor angle deviation keeps on fluctuating. Thus, the CCT with only SG present is 4.2s (6.2s-2s).

In the next simulation the Solar PV is introduced and connected to bus 4. It acts as a single centralized DG. The share by Solar PV is increased and when the share is 30MW voltage and maximum rotor angle deviation plots are shown in figure (5). It can be seen that transient stability of the system is greatly enhanced with the inclusion of Solar PV to the microgrid in addition to the existing SG. The new CCT is increased to 5.8s (7.8s-2s) from 4.2 of the case with only SG used.

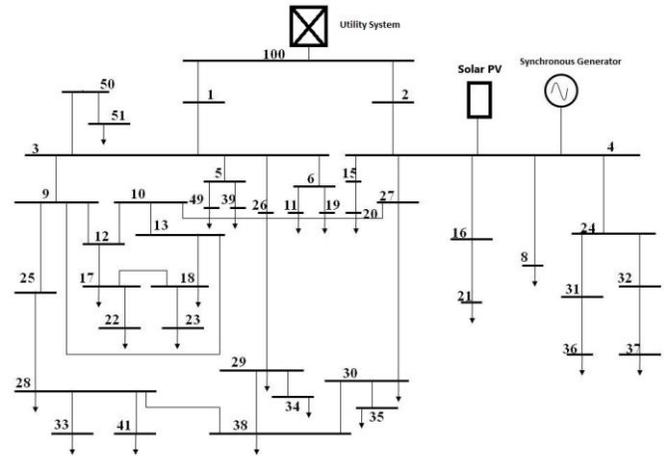

Fig. 2: IEEE Std 399-1997 test system

Several simulation is then performed by varying the share of Solar PV, and the change in CCT obtained is given in figure (6). Initially when the share by Solar PV is little, there is a drop in CCT compared to that of only SG used as DG (0 MW). When the share of solar PV is further increased the system gains more stability as CCT length becomes higher.

### C. With SG and multiple solar PVs distributed at random locations

This time simulation is performed in the presence of both SG and Solar PV. Instead of single PV unit multiple units are now used and the stability evaluated by measuring the CCT. At first 2 Solar PVs each of rating 2.5 MW are connected at bus 4 and 10. Then 2 PV units each of 5MW (total 10 MW) are connected at bus 4 & 10. Then 4 PV units each of 5MW (total 20 MW) are connected at bus 4,10,15 & 16 respectively and the new CCT measured. At next the procedure is followed for 5 PV units each of 5MW (total 25MW), connected to bus 4,10,15,16 and 27. Finally 6 PV units each comprising of 5MW (total 30MW) and connected to bus 4,8,10,15,16 and 27 are used to check stability.

A comparison is then made between multiple solar PVs distributed at random location and single large centralized Solar PV unit. The comparison is shown in bar format in the figure (7).

From figure (7) it can be stated that system stability is hampered when multiple Solar PVs are used as DG. Although the drop in CCT is minimum at lower rating generators but when share is high by PV the reduction in CCT is visible. When 30MW is supplied by PV source, CCT with multiple PVs is 5.3s, which is 0.5s less than that of a single unit of the same source.

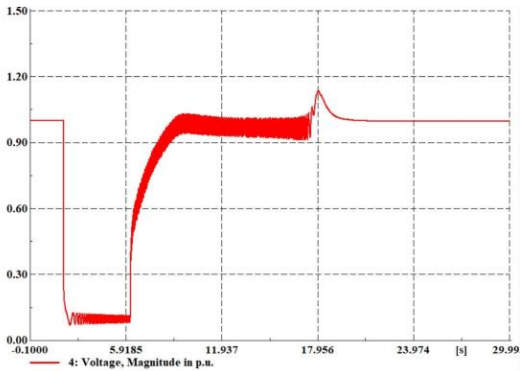 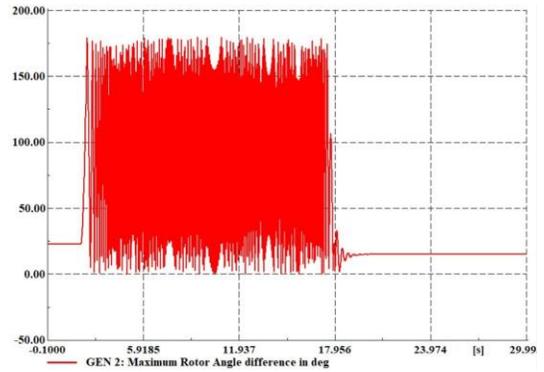

(a) Voltage curve in PU  (b) maximum rotor angle deviation in degrees

Fig. 3: Voltage and maximum rotor angle deviation plot with only SG & when fault cleared at 6.2s

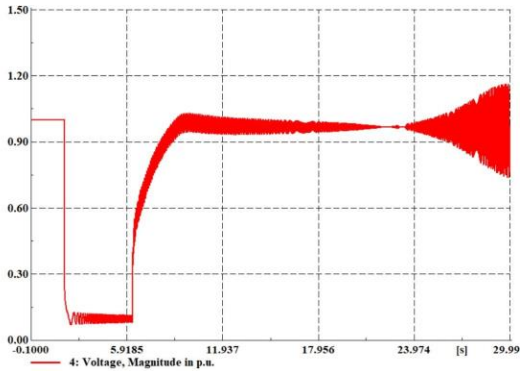 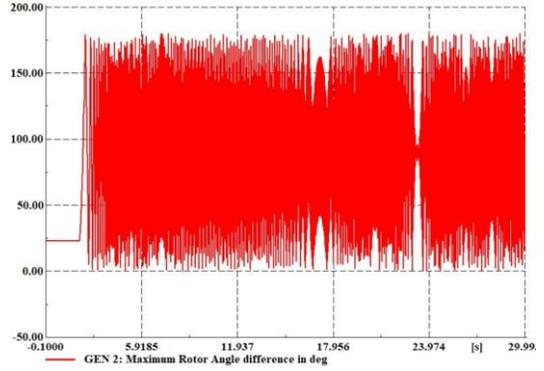

(a) Voltage curve in PU  (b) maximum rotor angle deviation in degrees

Fig. 4: Voltage and maximum rotor angle deviation plot with only SG & when fault cleared at 6.3s

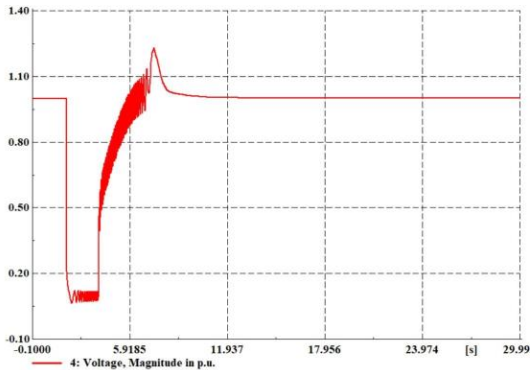 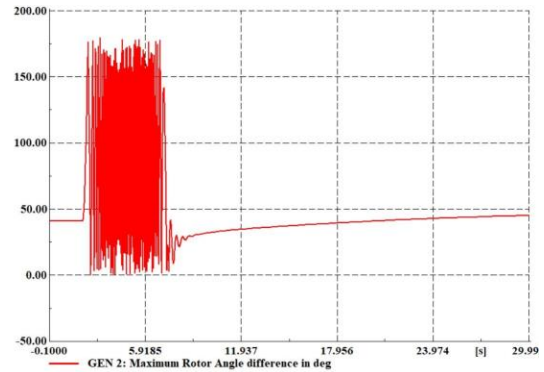

(a) Voltage curve in PU  (b) maximum rotor angle deviation in degrees

Fig. 5: Voltage and maximum rotor angle deviation plot with addition of centralized 30MW Solar PV & when fault cleared at 7.8s

In order to investigate the reason for such behavior the fault current was analyzed for both the scenarios as shown by figure (8). It was found that for the case of multiple generators the total combined 30MW generation resulted a 92 PU of fault current during short circuit fault. On the other hand, for the case of single PV unit the fault current is 83 PU. Thus, higher fault current resulted in the decrement of CCT and hence a reduction in the systems transient stability.

V. CONCLUSION

To make modern power system more reliable extensive study have been made and everyday new technologies are developed. Transient stability is the field that had been given the preference to improve systems stability. In most cases CCT is analyzed to check the systems response under faulty condition. Microgrid is rapidly growing in today's power system. During islanded operation certain challenges are faced when operating a microgrid. Industrial microgrids face further challenges due to presence of large number of induction motors.

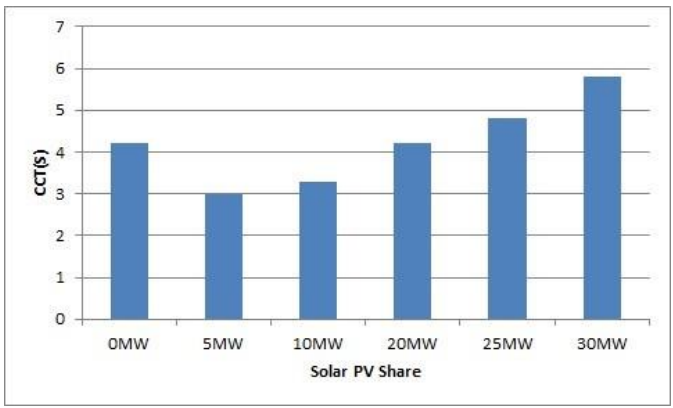

Fig. 6: Variation in CCT with change of single centralized Solar PV share

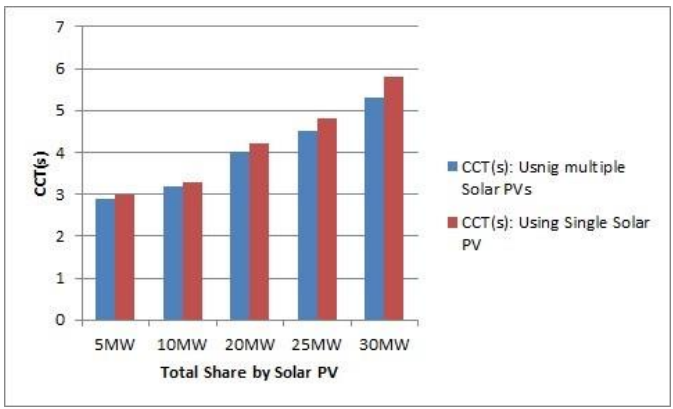

Fig. 7: Comparison of CCT between Single Centralized Solar PV and Multiple Solar PV units with total same rating

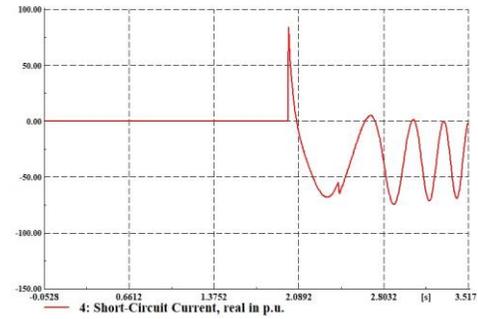

(a) with Single Centralized Solar PV

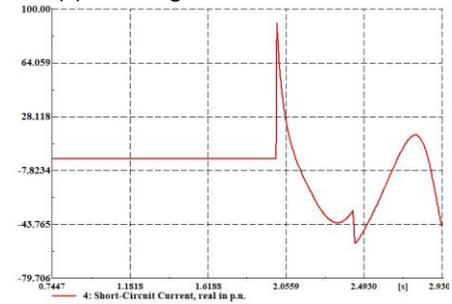

(b) With Multiple Solar PV

Fig. 8: Short circuit fault current at bus 4 for 2 different scenarios

This paper deals with the transient stability study of an industrial microgrid when solar PV is used as DG. Primary analysis was made to see what effect solar PV share have on the systems CCT. Results showed that when solar PV share is low CCT is lower than that of microgrid containing only SG. But as the share is increased CCT gradually increases and system gets more stable.

Then a comparative analysis was made to see the effects of using multiple solar PVs distributed throughout the network. Results verified that when multiple units are used fault current is increased. Thus, there is a drop in CCT of the system compared to the CCT of a microgrid containing large centralized single solar PV unit.

In future works the impact of change in insolation level of solar PV would be considered. Further the control aspects of the PV converters to reduce the fault current will be taken under consideration.